\documentclass[]{iau}
\usepackage{graphicx}
\usepackage[]{natbib}
\usepackage{aas_macros} 
\title[The Concerning S$H_0$ES Hubble Constant] 
{The Concerning S$H_0$ES Hubble Constant}

\author[]   
{Daniel Majaess}
\affiliation{Mount Saint Vincent University \\ Halifax, Canada \\ {\tt daniel.majaess@msvu.ca}}

\jname{The VVVX Survey: Exploitation \& Future IR Synoptic Survey}
\editors{Conference at the Vatican Observatory, Castel Gandolfo}
\begin{document}
\twocolumn[ 
\maketitle

\begin{abstract}
Concerns are raised regarding the S$H_0$ES results, and the present $H_0$ controversy.  The S$H_0$ES $H_0 \simeq 73$ km/s/Mpc has remained relatively unaltered across $18$ years (2005-2023), despite marked shifts in maser and Cepheid distances to the keystone galaxy NGC4258 (M106), and changes in the slope, zeropoint, metallicity, and extinction terms tied to the Leavitt Law, and notwithstanding uncertain photometry for remote Cepheids spanning galaxies with highly inhomogeneous crowding and surface brightness profiles.  Concerns raised regarding the S$H_0$ES findings by fellow researchers are likewise highlighted.  An independent blind assessment of the entire suite of raw HST Cepheid images is warranted, while being mindful of \textit{a priori} constraints and confirmation bias that unwittingly impact conclusions.   
\\ \\
Part of a talk given at the VVVX Survey conference, hosted at the Vatican Observatory, Castel Gandolfo (Oct.~2023).
\end{abstract}]


\section{Introduction}
The reputed `\textit{Hubble tension}' stems in part from an offset between the Planck CMB $H_0$ relative to that reported by the S$H_0$ES team.  The latter argue the difference separating the estimates is significant ($\simeq67$ vs.~$73$ km/s/Mpc), and the Planck team's determination is erroneous.  The S$H_0$ES team purports their conclusion is robust, and uncertainties associated with the Cepheid distance scale were mitigated \citep[e.g.,][]{rie16}.

\begin{figure}[t]
\begin{center}
 \includegraphics[width=3.2in]{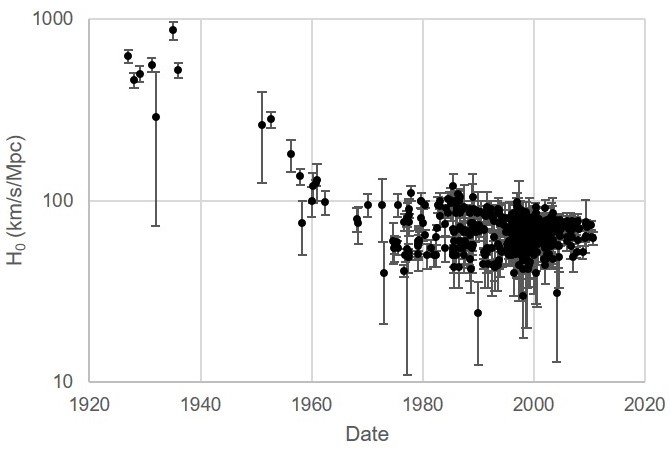} 
 \includegraphics[width=3.1in]{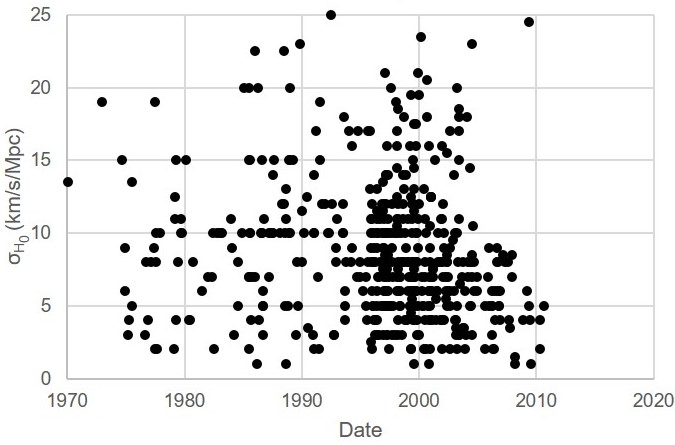}
 \caption{Historical $H_0$ estimates from the Huchra database.  Low uncertainties have been reported for decades.  Fig.~2 in \citet{ste20} features $H_0$ data beyond 2010.}
 \label{fighistorical}
\end{center}
\end{figure}

Present claims favoring a \textit{bona fide} offset between Cepheid and CMB $H_0$ determinations are premature given: a lack of consensus on the Leavitt Law (coefficients, zeropoint, and form), errors an anomalies endemic to S$H_0$ES findings, ongoing challenges to secure uncontaminated Cepheid photometry across remote spiral galaxies (e.g., NGC4258, \S \ref{sec-ngc4258}, Fig.~\ref{figngc4258}), a history of underestimated $H_0$ uncertainties (Fig.~\ref{fighistorical}), and owing to a spread in CMB results (Fig.~\ref{figcmb}).

\begin{figure*}[t]
\begin{center}
 \includegraphics[width=6.5in]{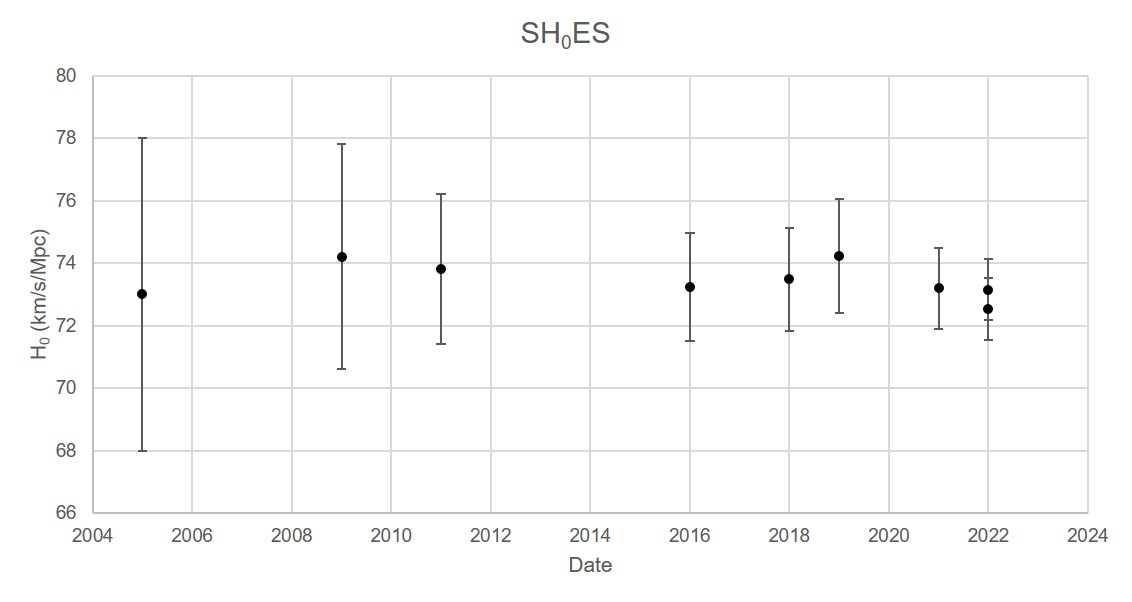} 
 \caption{A subsample of S$H_0$ES $H_0$ estimates across $\sim18$ years \citep[e.g.,][]{rie05}.  Their $H_0$ remained comparatively unchanged ($\simeq 73$ km/s/Mpc) despite ambiguities in the Leavitt Law, maser and $W_{VI}$ Cepheid distances for NGC4258, etc.~(\S \ref{sec-ngc4258}).}
 \label{figshoes}
\end{center}
\end{figure*}

\section{S$H_0$ES results \& NGC4258 Cepheids}
\label{sec-ngc4258}
 For example, Fig.~12 in \citet[][S$H_0$ES]{rie09b} implies that the slope ($\alpha$) of the classical Cepheid $W_{VI}$ function for near-solar Cepheids is $\alpha =-2.98\pm0.07$, as inferred mainly from SN-host galaxies (e.g., NGC3370). \citet{maj10} and \citet{maj11} countered that the $W_{VI}$ slope is comparatively constant across a sizable metallicity baseline ($\alpha \simeq -3.3$, $\Delta \rm{[Fe/H]} \simeq 1$), as established from Local Group Cepheids in the Magellanic Clouds, Milky Way, NGC6822, and IC1613 (Fig.~\ref{figmetallicity}).  The S$H_0$ES team ostensibly recanted their earlier position by citing a $W_{VI}$ slope tied to near-solar Cepheids of $\alpha \simeq -3.3$ \citep[][S$H_0$ES]{rie22}, whereas for example \citet[][S$H_0$ES]{rie09b} deduced $\alpha = -2.60\pm 0.24$ for NGC3021 Cepheids.

\begin{figure*}[t]
\begin{center}
 \includegraphics[width=6.9in]{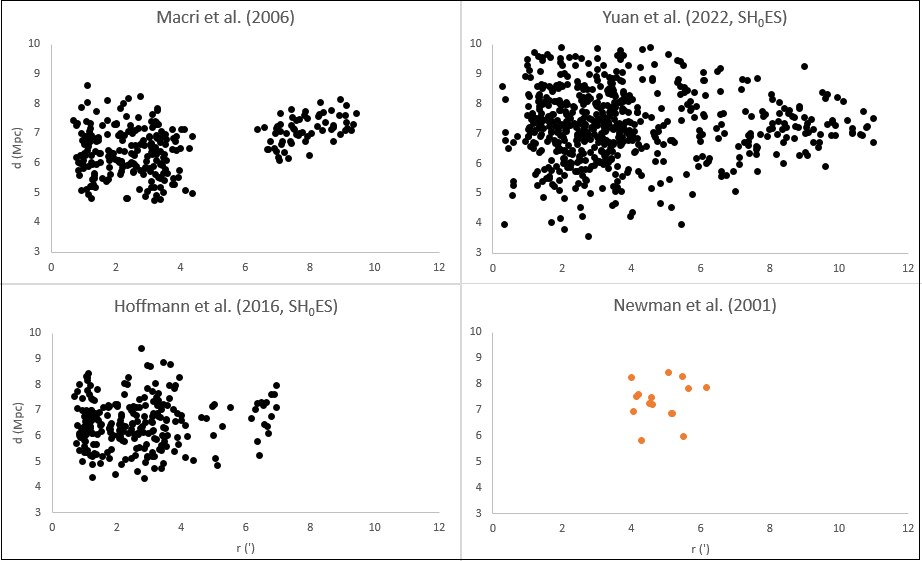} 
 \caption{NGC4258 Cepheid datasets exhibit striking differences with regards to:  their distances as a function of apparent galactocentric radius (x-axis), a (null-)dependence on abundance, and the mean distance.  Discrepancies emerge irrespective of whether a $W_{VI}$ Galactic Cepheid calibration \citep[][and references therein]{maj11} is used (shown), or the metallicity-dependent $W_{VI}$ relationship of \citet[][]{bre22}.} 
 \label{figngc4258}
\end{center}
\end{figure*}

\begin{figure}[t]
\begin{center}
 \includegraphics[width=3.4in]{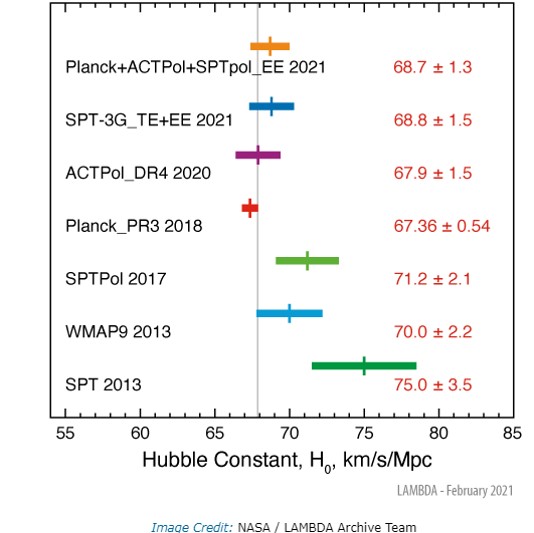} 
 \caption{$H_0$ estimates from diverse CMB surveys. There exist solutions beyond the Planck CMB, and which are comparable within the uncertainties to the \citet[][and references therein]{fm23b} TRGB analysis.}
 \label{figcmb}
\end{center}
\end{figure}

\begin{figure}[t]
\begin{center}
 \includegraphics[width=3.5in]{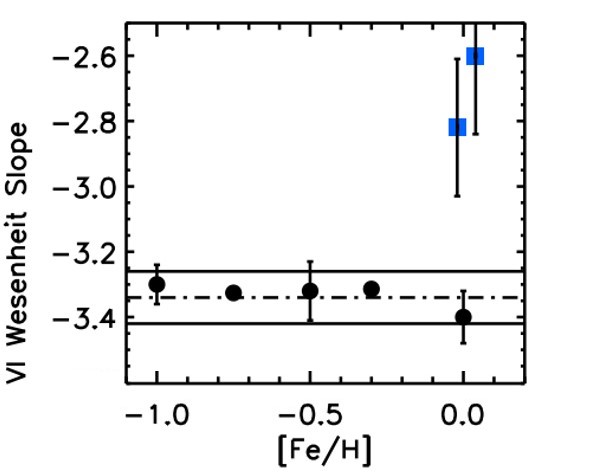} 
 \caption{\citet{maj10} indicated the $W_{VI}$ Leavitt Law's slope ($\alpha\simeq-3.3$) is insensitive to metallicity. \citet[][S$H_0$ES]{rie09b} argued that near-solar Cepheids adhered to $\alpha =-2.98\pm0.07$, a result they subsequently overturned \citep[$\alpha\simeq-3.3$,][S$H_0$ES]{rie22}. For example, \citet[][S$H_0$ES]{rie09b} determined that NGC3021 Cepheids follow $\alpha =-2.60\pm 0.24$.}
 \label{figmetallicity}
\end{center}
\end{figure}

The history of Cepheid and maser distances to NGC4258 is disconcerting. \citet{mao99} assessed HST images to determine a Cepheid distance for NGC4258 of $8.1\pm0.4$ Mpc, which was discrepant relative to the \citet{her99} maser distance of $7.2\pm0.3$ Mpc. The \citet{mao99} team revisited their Cepheid distance in \citet{new01}, and revised it downward ($7.8\pm0.3\pm0.5$ Mpc) and marginally closer to the \citet{her99} maser result.  At the time various scenarios were proposed aiming to reconcile the two estimates \citep[e.g.,][]{new01,cap02}. \citet{mac06} analyzed HST images of Cepheids in two separate NGC4258 fields, and discovered a substantial distance offset between them ($0^{\rm m}.16$), which is readily discernible in Fig.~\ref{figngc4258}, and that was established in this instance using a Galactic $W_{VI}$ calibration \citep[][and references therein]{maj11}.  Importantly, an insidious degeneracy exists whereby more crowded fields at smaller galactocentric radii are comparatively metal-rich, whereas increasingly metal-poor Cepheids extend outward to the galaxy's lower surface brightness periphery.  That degeneracy can compromise determinations of the impact of chemical composition on Cepheid distances \citep[e.g.,][their \S 5]{Mac01}.  Indeed, the lack of consensus on the Leavitt Law partly stems from the aforementioned degeneracy.  \citet{maj12b,maj12a,maj16} noted that photometric contamination was likewise problematic for certain globular clusters and the Galactic Center sightline.\footnote{Paradoxically, \citet{maj20} confirmed that blending with say red clump giants may advantageously thrust otherwise faint extragalactic targets (e.g., RR Lyrae variables) into the range of detectability \citep[see also \S 2.4 of][]{maj18}.}   \citet[][]{yua20} identified that Cepheids in high surface brightness regions of the Seyfert 1 galaxy NGC4151 exhibited a systematic shift (their Fig.~9, right panel).  Rather than favoring the contamination scenario, \citet{mac06} attributed their NGC4258 variations to a Leavitt Law zeropoint that's metallicity dependent ($\gamma=-0.29\pm0.09\pm0.05$ mag/dex).  However, \citet{maj10} and \citet{bre11} argued that the abundance gradient across NGC4258 may be shallower\footnote{Table~12 in \citet{rie09b} and Fig.~4 in \cite{maj10}.} than that finally selected by \citet[][\S 4.3]{mac06}, which would markedly expand their cited Cepheid metallicity effect to an unrealistic value.  Yet there are those favoring an extreme Cepheid metallicity effect, such as \citet[][$-0.80\pm0.21$ mag/dex]{sha11} and \citet[][$-0.61\pm0.21$ mag/dex]{fau15}.  Critically, \citet{maj11} disagreed with those conclusions by relaying that applying such an immense metallicity dependence yielded spurious Cepheid distances for the Magellanic Clouds (e.g., $\mu_{0,LMC}\neq18.1$). \citet{uda01}, \citet{maj11}, \citet{wie17}, and \citet{mf23} concluded that $W_{VI}$ functions are relatively insensitive to metallicity,\footnote{\citet{bon08} and \citet{and16} disagree regarding what models produced relative to the Leavitt Law's dependence on metallicity.  \S 5.5 of \citet[][]{bre22} presents their viewpoint concerning the \citet{wie17} result.} whereas \citet[][S$H_0$ES]{rb23} advocate for a larger zeropoint dependence by comparison (e.g., $-0.22\pm0.04$ mag/dex).  \citet{mf23} relayed TRGB-Cepheid distances (their Fig.~1) which overturn the existing \citet{sak04} analysis, and likewise contest \citet[][$-0.201\pm0.071$ mag/dex]{bre22}. Moreover, the reader is encouraged to examine Fig.~6 in \citet[][S$H_0$ES]{yua22} where a constant (indicating a null-dependence) can represent their latest NGC4258 analysis rather than the fits they overlaid.  \citet[][S$H_0$ES]{yua22} constrained the dependence to $-0.07\pm0.21$ mag/dex, which together with their overall data are in stark contrast to the \citet{mac06} and \citet[][S$H_0$ES]{Hof16} interpretations of NGC4258, and the reader can arrive at their own conclusion by inspecting Fig.~\ref{figngc4258} \citep[see also \S4 in][]{yua22}.  Alarmingly, there's a significant $I$-band (F814W) and $W_{VI}$ discrepancy between \citet[][S$H_0$ES]{Hof16} and \citet[][S$H_0$ES]{yua22}, which is characterized by a considerable mean difference ($\gtrsim0^{\rm m}.15$).  Regardless of applying a Galactic $W_{VI}$ Cepheid calibration \citep{ben07,maj11} or the \citet[][]{bre22} metallicity-dependent $W_{VI}$ function to the NGC4258 datasets of \citet{mao99}, \citet{new01}, \citet{mac06}, \citet{fau15}, \citet[][S$H_0$ES]{Hof16}, and \citet[][S$H_0$ES]{yua22}: discrepant results arise.  

Offsets in Cepheid distances across remote galaxies possessing non-uniform crowding and surface brightness are comparatively unassociated with a $W_{VI}$ metallicity effect, but rather point to serious shortcomings in establishing homogeneous  uncontaminated photometry, as echoed previously \citep[e.g.,][]{maj11}.  The offset highlighted by \citet[][their \S 5]{Mac01} resides elsewhere in extragalactic Cepheid data.  

The \citet[][$7.2\pm0.3$ Mpc]{her99} maser distance for NGC4258 was superseded by \citet[][$7.60\pm0.17\pm0.15$ Mpc, see also \S 3 in \citealt{rie16}]{hum13}.  \citet{rei19} subsequently determined $d=7.576\pm0.082\pm0.076$ Mpc.  

\section{Independent Cepheid-TRGB Analyses}
S$H_0$ES-independent Cepheid and TRGB based analyses include the HST key project to measure $H_0$, namely $68\pm5$ and $72\pm8$ km/s/Mpc \citep{gib00,fre01}. Moreover, \citet{tam13} cite a final determination of $H_0=64.1\pm2.0$ km/s/Mpc \citep[see also][]{ts10}.  The `\textit{Hubble tension'} debate emerged because of the latter's passing.  Admittedly, \citet{maj10} and others argued that the \citet{san04} distance scale was too remote \citep[e.g.,][]{ben07,rie09b}, and hence their $H_0$ was underestimated.  However, the impact of photometric contamination was not comprehensively addressed by that team and would shift $H_0$ in the opposite direction.  

Concerns emerged near the conclusion of the HST key project to secure $H_0$ that extraneous flux from stars along the Cepheid's sightline could lead to an overestimated expansion rate \citep[e.g.,][]{su99}, and the reader is encouraged to review the debate and rebuttals in \S 8.5 of \citet{fre01}, \S 7 in \citet{moc00}, and \S 8 in \citet{moc01}.  \citet{fre01} attributed a sizable uncertainty to the phenomenon ($0^{\rm m}.1$), and the \citet[][CHP]{fre12} Cepheid effort shouldn't be leveraged to support the S$H_0$ES Cepheid $H_0$ since it relies on the earlier potentially contaminated HST key project data (c2000).\footnote{\citet{fre12} advance a separate position in their \S3.3 and appendix.}  Importantly, the current \citet{fm23b} TRGB efforts are pertinent in part because the team's \textit{modus operandi} is to avoid crowded regions. Their $H_0$ is comparable to diverse CMB measurements (Fig.~\ref{figcmb}).  Indeed, a consensus framework on the topic of photometric contamination remains outstanding, in tandem with the terminology employed \citep[][\S4 in \citealt{yua22}, and see also the disparate discussions in \citealt{rb23} and \citealt{fm23}]{maj20}.  The reader can promptly grasp the difficulties faced by inspecting Fig.~1 in \citet{moc01}, Fig.~2 in \citet{maj12a}, and critically Fig.~8 in \citet{fm23}, which conveys HST crowding relative to JWST.

\section{Conclusion}
Historical $H_0$ uncertainties (Fig.~\ref{fighistorical}), errors and anomalies existing within S$H_0$ES and NGC4258 data (\S \ref{sec-ngc4258}), a relatively unchanged S$H_0$ES $H_0$ across $\sim18$ years (Fig.~\ref{figshoes}), diversity among CMB measurements (Fig.~\ref{figcmb}), and independent results from \citet{mf23} and \citet{fm23b,fm23}: indicate in concert that claims of a \textit{bona fide} offset between Cepheid and CMB $H_0$ findings should be viewed cautiously. 

 The aforementioned ambiguities continue a long established trend tied to $H_0$ research (Fig.~\ref{fighistorical}), from say Sandage and de Vaucouleurs \citep{ove91}, to \citet{fre01} and \citet{san04}, and now between \citet[][S$H_0$ES]{rie16}, \citet{fm23b}, and others.  TRGB and Cepheid research by groups and (co)authors beyond S$H_0$ES and (C)CHP are desirable, and that includes a reassessment of the entire raw HST Cepheid imagery without biased constraints that inadvertently sway conclusions. The case of NGC4258 broadly illustrates that a chief concern regarding $H_0$ determinations is the  challenging task of obtaining precise, commonly standardized, multiepoch, multiband, uncontaminated remote extragalactic Cepheid photometry \citep[e.g.,][]{maj10}.  

 Continued research is needed to assess Gaia's data.  The spurious Hipparcos distances to the Pleiades and Blanco 1 are key lessons pointing toward enhanced vigilance \citep{maj11b}.  Moreover, oft-cited NGC4258 and LMC distances could likewise be incorrect, and certain results may require adjustment that conspire to sway $H_0$ unidirectionally.  Lastly, challenging degeneracies not only exist between blending and characterizing the effect of metallicity on Cepheid distances, but possibly also with respect to non-standard extinction across a galaxy \citep[e.g.,][]{tur13,car13}. On that note a discrepant Cepheid distance exists for NGC5128 or Cen A \citep[][]{fer07,maj10,har10}, and a non-canonical extinction law has been debated in that case (see also \citealt{fau15} regarding NGC4258).

\section*{Acknowledgements}
\normalsize{D.M. is grateful to the conference's fellow participants, and the welcoming residents of Castel Gandolfo and Albano. This research relies on the efforts of the following initiatives: CDS, NASA ADS, arXiv, OGLE, Araucaria, S$H_0$ES, (C)CHP.}

\bibliographystyle{apjMOD} 
\bibliography{talk} 
\end{document}